\documentclass[aps,twocolumn]{revtex4}
\usepackage[dvips]{graphicx}
\usepackage{color}

\begin{document}

\title{Connectivity and expression in protein networks: \\
Proteins in a complex are uniformly expressed}


\author{Shai Carmi$^1$, Erez Y. Levanon$^2$, Shlomo Havlin$^1$, Eli Eisenberg$^3$}

\affiliation{$^1$Minerva Center and Dept.\ of Physics,
Bar-Ilan University, Ramat-Gan 52900, Israel}
\affiliation{$^2$Compugen Ltd., 72 Pinhas Rosen St., Tel-Aviv 69512, Israel}
\affiliation{$^3$School of Physics and Astronomy, Raymond and Beverly Sackler
Faculty of Exact Sciences, Tel Aviv University, Tel Aviv 69978, Israel}

\begin{abstract}
We explore the interplay between the protein-protein interactions network
and the expression of the interacting proteins.
It is shown that interacting proteins are expressed in
significantly more similar cellular concentrations. This is
largely due to interacting pairs which are part of protein
complexes. We solve a generic model of complex formation and show
explicitly that complexes form most efficiently when their
members have roughly the same concentrations. Therefore, the
observed similarity in interacting protein concentrations could be
attributed to optimization for efficiency of complex formation.
\end{abstract}


\maketitle

\section{Introduction}

Statistical analysis of real-world networks topology has attracted
much interest in recent years, proving to supply new insights and
ideas to many diverse fields. In particular, the protein-protein
interaction network, combining many different interactions of
proteins within a cell, has been the subject of many studies
(for a recent review see \cite{Barabasi}).
While this network shares many of the universal
features of natural networks such as the scale-free distribution
of degrees \cite{Jeong}, and the small world characteristics
\cite{flag}, it also has some unique features. One of the most
important of these is arguably the fact that the protein interactions
underlying this network can be separated into two roughly disjoint
classes. One of them relates to transmission of information within
the cell: protein A interacts with protein B and changes it, by a
conformational or chemical transformation. The usual scenario
after such an interaction is that the two proteins disassociate
shortly after the completion of the transformation. On the other
hand, many protein interactions are aimed at the formation of a
protein complex. In this mode of operation the physical attachment
of two or more proteins is needed in order to allow for the
biological activity of the combined complex, and is typically
stable over relatively long time scales \cite{Han}.

The yeast {\it Saccharomyces cerevisiae} serves as the model
organism for most of the analyses of protein-protein interaction
network. The complete set of genes and proteins with extensive
data on gene expression are available \cite{Cherry} for this
unicellular organism, accompanied by large datasets of
protein-protein interactions based on a wide range of experimental
and computational methods
\cite{synexp,mrna,genefus,hms,y2h,synleth,2nei,tap,von Mering}.
In addition, the intracellular locations
and the expression levels of most proteins of the yeast were recently
reported \cite{Ghaemmaghami}. The availability of such data enables
us to study the relationship between network topology and the
expression levels of each protein.

In this work we demonstrate the importance of the distinction
between different types of protein interaction, by highlighting
one property which is unique to interactions of the protein
complexes. Combining databases of yeast protein interactions
with the recently reported information on the protein
concentration, we find that proteins belonging to the same complex
tend to have a more uniform concentration distribution. We further
explain this finding by a model of complex formation, showing that
uneven concentrations of the complex members result in inefficient
complex formation. Surprisingly, in some cases increasing the
concentration of one of the complex ingredients decreases the
absolute number of complexes formed. Thus, the experimental
observation of uniform complex members concentrations can be
explained in terms of selection for efficiency.

\section{Concentrations of Interacting Proteins}

\begin{figure}[htb]
\centering
\includegraphics[totalheight=1.5in]{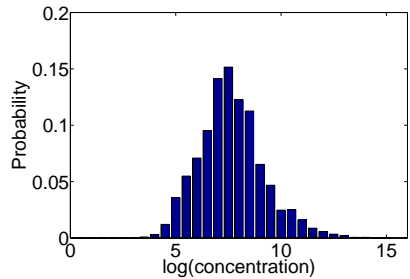}
\caption{ {(Color online) Distribution of the logarithm of the protein
concentration (in units of protein molecules per cell) for all measured 
proteins within the yeast cell.}}
\label{lognormal}
\end{figure}

We start by studying the concentrations of pairs of interacting proteins,
and demonstrate that different types of protein-protein interactions differ
in their properties. For this purpose we use the recently published database
providing the (average) concentration \cite{Ghaemmaghami},
as well as the localization within the cell,
for most of the {\it Saccharomyces cerevisiae}
(baker's yeast) proteins \cite{Huh}.
The concentrations $c_i$ (given in arbitrary units)
are approximately distributed according to a log-normal distribution
with $\langle log(c_i)\rangle=7.89$
and standard deviation $1.53$ (Fig. \ref{lognormal}).

The bakers' yeast serves as a model organism for most of the
protein-protein interaction network studies. Thus a set of many of
its protein-protein interactions is also readily available. Here
we use a dataset of recorded yeast protein interactions, given
with various levels of confidence \cite{von Mering}. The dataset
lists about $80000$ interactions between approximately $5300$
of the yeast proteins (or about $12000$ interactions between
$2600$ proteins when excluding interactions of the lowest
confidence). These interactions were deduced by many different
experimental methods, and describe different biological relations
between the proteins involved. The protein interaction network
exhibits a high level of clustering (clustering coefficient
$\approx0.39$). This is partly due to the existence of many sets
of proteins forming complexes, where each of the complex members
interacts with many other members.

Combining these two databases, we study the correlation between the
(logarithm of) concentrations of pairs of interacting
proteins. In order to gain insight into the different components
of the network, we perform this calculation separately for the interactions
deduced by different experimental methods.
For simplicity, we report here the results after excluding the interactions
annotated as
low-confidence (many of which are expected to be false-positives).
We have explicitly checked that their inclusion does not change the results
qualitatively. The results are summarized in Table \ref{table}, and show a
significant correlation between the expression levels of
interacting proteins.

\begin{widetext}
\begin{table}[b]
\begin{tabular}{|*{8}{c|}}
\hline Interaction & Number of & Number of & Number
of & Correlation & STD of & P-Value \\
& interacting & interactions & interactions in &
between & random &  \\
& proteins &&  which expression & expression &  correlations \\
&&& level is & levels of &&   \\
&&&  known for & interacting &&  \\
&&&  both proteins &     proteins && \\
\hline
All & 2617 & 11855 & 6347 & 0.167 & 0.012 & $10^{-42}$ \\
\hline
Synexpression\cite{synexp,mrna}
& 260 & 372 & 200 & 0.4 & 0.065 & $3.5\cdot10^{-10}$ \\
\hline
Gene Fusion\cite{genefus} & 293 & 358 & 174 & -0.079 & - & - \\
\hline
HMS\cite{hms} & 670 & 1958 & 1230 & 0.164 & 0.027 & $3.3\cdot10^{-10}$ \\
\hline
yeast 2-Hybrid\cite{y2h}& 954 & 907 & 501 & 0.097 & 0.046 & $1.7 \cdot 10^{-2}$ \\
\hline
Synthetic Lethality\cite{synleth} & 678 & 886 & 497 & 0.285 & 0.045 & $1.2\cdot10^{-10}$ \\
\hline
2-neighborhood\cite{2nei} & 998 & 6387 & 3110 & 0.054 & 0.016 & $5.4\cdot10^{-4}$ \\
\hline
TAP\cite{tap} & 806 & 3676 & 2239 & 0.291 & 0.02 & $10^{-49}$ \\
\hline
\end{tabular}
\caption{ (Color online) Correlation coefficients between the logarithm of the
concentrations of interacting proteins. Only interactions of medium
or high confidence were included. The statistical
significance of the results was estimated by randomly permuting
the concentrations of the proteins and reevaluating the
correlation on the same underlying network,  repeated for 1,000
different permutations. The mean correlation of the randomly
permuted networks was zero, and the standard deviation (STD) is given.
The P-value was calculated assuming gaussian distribution of the
correlation values for the randomized networks. We have verified
that the distributions of the 1,000 realizations calculated are
roughly Guassian.} \label{table}
\end{table}
\end{widetext}

The strongest correlation is seen for the subset of protein
interactions which were derived from synexpression, i.e. inferred
from correlated mRNA expression. This result confirms the common
expectation that genes with correlated mRNA expression would yield
correlated protein levels as well\cite{mrna}. However, our results show that
interacting protein pairs whose interaction was deduced by other methods
exhibit significant positive correlation as well. The effect is weak for
the yeast 2-Hybrid (Y2H) method\cite{y2h} which includes all possible
physical interactions between the proteins (and is also known to
suffer from many artifacts and false-positives), but stronger for
the HMS (High-throughput Mass Spectrometry)\cite{hms} and TAP
(Tandem-Affinity Purification)\cite{tap} interactions corresponding to
actual physical interactions (i.e., experimental evidence that the
proteins actually bind together in-vivo). These experimental
methods are specifically designed to detect cellular protein
complexes. The above results thus hint that the overall
correlation between concentrations of interacting proteins is due
to the tendency of proteins which are part of a stable complex to
have similar concentrations.

The same picture emerges when one counts the number of
interactions a protein has with other proteins of similar
concentration, compared to the number of interactions with
randomly chosen proteins. A protein interacts, on average, with
$0.49\%$ of the proteins with similar expression level (i.e.,
$|$log-difference$| < 1$), as opposed to only $0.36\pm 0.01~\%$ of
random proteins, in agreement with the above observation of
complex members having similar protein concentrations.

In order to directly test this hypothesis (i.e. that proteins in a
complex have similar concentrations), we use existing datasets of
protein complexes and study the uniformity of concentrations of
members of each complex. The complexes data were taken from
\cite{complexes}, and were found to have many TAP interactions
within them.
As a measure of
the uniformity of the expression levels within each complex, we
calculate the variance of the (logarithm of the)
concentrations among the members of each complex.
The average variance (over all complexes) is found to be
$2.35$, compared to $2.88\pm0.07$ and $2.74\pm0.11$
for randomized complexes in two different randomization schemes (see
figure), confirming that the concentrations of complex members
tend to be more uniform than a random set of proteins.

\begin{figure}[htb]
\centering
\includegraphics[totalheight=1.5in]{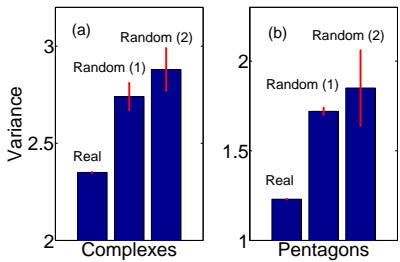}
\caption{{(Color online) (a) Variance of the logarithm of the
protein expression levels (in units of mulecules per cell)
for members of real complexes, averaged over all complexes,
comapred with the averaged variance of the complexes after
randomization of their members, letting each protein participate on average in
the same number of complexes (random(1)),
as well as randomized complexes where the number
of complexes each protein participates in is kept fixed (random(2)).
Real complexes have a lower variance, indicating higher uniformity in the
expression levels of the underlying proteins. (b) Same as (a) for
expression levels in pentagons (see text).}} \label{unity}
\end{figure}

As another test, we study a different yeast protein interaction
network, the one from the DIP database \cite{Xenarios}. We look
for fully-connected sub-graphs of size $5$, which are expected to
represent complexes, sub-complexes or groups of proteins working
together. The network contains approximately $1600$ (highly overlapping)
such pentagons, made of about $300$ different proteins. The
variance of the logarithm of the concentrations of each
pentagon members, averaged over the different pentagons, is 1.234.
As before, this is a significantly low variance compared with random
sets of five proteins (average variance $1.847 \pm 0.02$ and $1.718\pm0.21$),
see figure \ref{unity}.

Finally, we have used mRNA expression data \cite{mrna} and looked
for correlated expression patterns within complexes. We have
calculated the correlation coefficient between the expression data
of the two proteins for each pair of proteins which are part of
the same pentagon. The average correlation coefficient between
proteins belonging to the same fully-connected pentagon is $0.15$
compared to $0.056\pm0.005$ for a random pair.

In summary, combination of a number of yeast protein interaction
networks with protein and mRNA expression data yields the
conclusion that interacting proteins tend to have similar
concentrations. The effect is stronger when focusing on
interactions which represent stable physical interactions, i.e.
complex formation, suggesting that the overall effect is largely
due to the uniformity in the concentrations of proteins belonging
to the same complex. In the next Section we explain this finding
by a model of complex formation. We show, on general grounds, that
complex formation is more effective when the concentrations of its
constituents is roughly the same. Thus, the observation made in
the present Section can be explained by selection for efficiency
of complex formation.

\section {Model}
Here we study a model of complex formation, and explore the
effectiveness of complex production as a function of the relative abundances of
its constituents. For simplicity, we start by a detailed analysis of the
three-components complex production, which already captures most
of the important effects.

Denote the concentrations of the three components of the complex
by $A$, $B$ and $C$, and the concentrations
of the complexes they form by $AB$, $AC$, $BC$ and $ABC$.
The latter is the concentration of the full complex, which is
the desired outcome of
the production, while the first three describe the different sub-complexes
which are formed (in this case, each of which is composed of two components).
Three-body processes, i.e., direct generation
(or decomposition) of $ABC$ out of $A$ $B$ and $C$, can usually be neglected
\cite{book}, but their inclusion here does not complicate the analysis.
The resulting set of reaction kinetic equations is given by

\begin{widetext}
\begin{eqnarray}
\frac{d(A)}{dt} & = & k_{d_{A,B}} AB + k_{d_{A,C}} AC +
(k_{d_{A,BC}} + k_{d_{A,B,C}}) \cdot ABC \nonumber\\ && - k_{a_{A,B}} A
\cdot B - k_{a_{A,C}} A \cdot C - k_{a_{A,BC}} A \cdot BC -
k_{a_{A,B,C}} A \cdot B \cdot C \\
\frac{d(B)}{dt}& = & k_{d_{A,B}} AB + k_{d_{B,C}} BC +
(k_{d_{B,AC}} + k_{d_{A,B,C}}) \cdot ABC \nonumber\\ &&- k_{a_{A,B}} A
\cdot B - k_{a_{B,C}} B \cdot C - k_{a_{B,AC}} B \cdot AC - k_{a_{A,B,C}} A \cdot B \cdot C\\
\frac{d(C)}{dt}& = & k_{d_{A,C}} AC + k_{d_{B,C}} BC +
(k_{d_{C,AB}} + k_{d_{A,B,C}}) \cdot ABC \nonumber\\ && - k_{a_{A,C}} A
\cdot C - k_{a_{B,C}} B \cdot C -
k_{a_{C,AB}} C \cdot AB - k_{a_{A,B,C}} A \cdot B \cdot C\\
\frac{d(AB)}{dt} & = & k_{a_{A,B}} A \cdot B + k_{d_{C,AB}} ABC -
k_{d_{A,B}} AB - k_{a_{C,AB}} C \cdot AB\\
\frac{d(AC)}{dt} & = & k_{a_{A,C}} A \cdot C + k_{d_{B,AC}} ABC -
k_{d_{A,C}} AC - k_{a_{B,AC}} B \cdot AC\\
\frac{d(BC)}{dt} & = & k_{a_{B,C}} B \cdot C + k_{d_{A,BC}} ABC -
k_{d_{B,C}} BC - k_{a_{A,BC}} A \cdot BC\\
\frac{d(ABC)}{dt} & = & k_{a_{A,BC}} A \cdot BC + k_{a_{B,AC}} B
\cdot AC + k_{a_{C,AB}} C \cdot AB + k_{a_{A,B,C}} A \cdot B \cdot
C \nonumber\\ && - (k_{d_{A,BC}} + k_{d_{B,AC}} + k_{d_{C,AB}} +
k_{d_{A,B,C}}) \cdot ABC
\end{eqnarray}
\end{widetext}
where $k_{a_{x,y}}$ ($k_{d_{x,y}}$) are the association
(dissociation) rates of the subcomponents $x$ and $y$ to form the
complex $xy$. Denoting the total number of type $A$, $B$ and $C$
particles by $A_0$, $B_0$, $C_0$, respectively, we may write the
conservation of material equations:
\begin{eqnarray}
A + AB + AC + ABC = A_0\\
B + BC + AB + ABC = B_0\\
A + AC + BC + ABC = C_0
\end{eqnarray}

We look for the steady-state solution of these equations, where
all time derivatives vanish. For simplicity, we consider first the
totally symmetric situation, where all the ratios of association
coefficients to their corresponding dissociation coefficients are equal,
i.e., the ratios $k_{d_{x,y}}/k_{a_{x,y}}$ are all equal to
$X_0$ and $k_{d_{x,y,z}}/k_{a_{x,y,z}}=X_0^2$, where $X_0$ is a constant with 
concentrations units. In this case,
measuring all concentrations in units of $X_0$, all the
reaction equations are solved by the substitutions $AB = A \cdot
B$, $AC = A \cdot C$, $BC = B \cdot C$ and $ABC = A \cdot B \cdot
C$, and one needs only to solve the material conservation
equations, which take the form:
\begin{eqnarray}
A + A \cdot B + A \cdot C + A \cdot B \cdot C = A_0\\
B + B \cdot C + A \cdot B + A \cdot B \cdot C = B_0\\
A + A \cdot C + B \cdot C + A \cdot B \cdot C = C_0
\end{eqnarray}
These equations allow for an exact and straight-forward (albeit
cumbersome) analytical solution. In the following, we explore the
properties of this solution. The efficiency of the production of
$ABC$, the desired complex, can be measured by the number of
formed complexes relative to the maximal number of complexes
possible given the initial concentrations of supplied particles
${\rm eff} \equiv ABC / \min{(A_0,B_0,C_0)}$. This definition does
not take into account the obvious waste resulting from proteins of
the more abundant species which are bound to be leftover due to
shortage of proteins of the other species. In the following we
show that having unmatched concentrations of the different complex
components result in lower efficiency beyond this obvious waste.

\begin{figure}[htb]
\centering
\includegraphics[totalheight=1.5in]{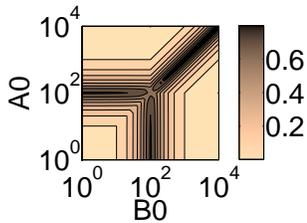}
\caption{{(Color online) The efficiency of the synthesis ${\rm eff} \equiv
ABC / \min{(A_0,B_0,C_0)}$ as a function of $A_0$ and $B_0$, for
$C_0=10^2$. The efficiency is  maximized when the two most
abundant species have roughly the same concentration.}} \label{1k}
\end{figure}

In the linear regime, $A_0, B_0, C_0 \ll 1$, the fraction of
particles forming complexes is small, and all concentrations are
just proportional to the initial concentrations. The overall
efficiency of the process in this regime is extremely low,
$ABC=A\cdot B\cdot C \sim A_0\cdot B_0\cdot C_0\ll A_0,B_0,C_0$.
We thus go beyond this trivial linear regime, and focus on the
region where all concentrations are greater than unity. Fig.
\ref{1k} presents the efficiency as a function of $A_0$ and $B_0$,
for fixed $C_0 = 10^2$. The efficiency is maximized when the two
more abundant components have approximately the same concentration,
i.e., for $A_0 \approx B_0$ (if $C_0<A_0,B_0$), for $A_0\approx C_0=10^2$
(if $B_0<A_0,C_0$) and for $B_0\approx C_0=10^2$
(if $A_0<B_0,C_0$).

\begin{figure}[htb]
\centering
\includegraphics[totalheight=1.5in]{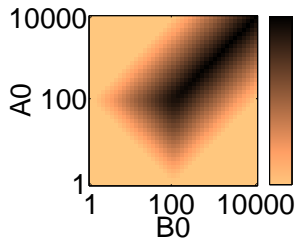}
\caption{(Color online) {$\log{(ABC)}$ as a function of $A_0,B_0$, for fixed
$C_0 = 10^2$.
For each row (fixed $A_0$) or column (fixed $B_0$) in the graph,
$ABC$ has a maximum, which occurs where $A_{0,max}
\approx\max{(B_0,C_0)}$ (for columns), and $B_{0,max}
\approx\max{(A_0,C_0)}$ (for rows).}} \label{a0max}
\end{figure}

Moreover, looking at the absolute quantity of the complex product,
one observes (fixing the concentrations of two of substances,
e.g., $B_0$ and $C_0$) that $ABC$ itself has a maximum at some
finite $A_0$, i.e., there is a finite optimal concentration for
$A$ particles (see Fig. \ref{a0max}). Adding more molecules of
type $A$ beyond the optimal concentration {\it decreases} the
amount of the desired complexes. The concentration that maximizes
the overall production of the three-component complex is
$A_{0,max} \approx \max{(B_0,C_0)}$.

\begin{figure}[htb]
\centering
\includegraphics[totalheight=2.5in]{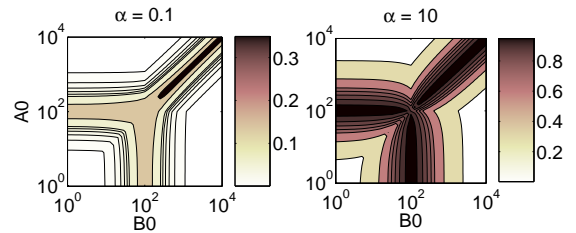}
\caption{(Color online) {Synthesis efficiency ${\rm eff}\equiv ABC /
\min{(A_0,B_0,C_0)}$ as a function of $A_0$ and $B_0$, for
different values of $\alpha$. $C_0$ is fixed, $C_0=100$. The
efficiency is maximized when the two most abundant substances are
of roughly the same concentration, regardless of the values of
$\alpha$.}} \label{4k}
\end{figure}

An analytical solution is available for a somewhat more general situation,
allowing the ratios $k_{d_{x,y}}/k_{a_{x,y}}$ to take different
values for the two-components association/dissociation ($X_0$) and
the three-components association/dissociation ($X_0/\alpha$ and 
$X_0^2/\alpha$ for association/dissociation of the three-component complex
from/to a two-component complex plus one single particle or to three 
single particles, respectively).
It can be easily seen that under these conditions, and measuring the
concentration in units of $X_0$ again,
the solution of the reaction kinetics equations is given by
\begin{eqnarray}
AB & = & A \cdot B, \\
AC & = & A \cdot C, \\
BC & = & B \cdot C, \\
ABC & = & \alpha ~ A \cdot B \cdot C,
\end{eqnarray}
and therefore the conservation of material equations take the form
\begin{eqnarray}
A + A \cdot B + A \cdot C + \alpha A \cdot B \cdot C = A_0\\
B + B \cdot C + A \cdot B + \alpha A \cdot B \cdot C = B_0\\
A + A \cdot C + B \cdot C + \alpha A \cdot B \cdot C = C_0
\end{eqnarray}
These equations are also amenable for an analytical solution, and
one finds that taking $\alpha$ not equal to $1$
does not qualitatively change the above results. In particular,
the synthesis is most efficient when the two highest concentrations are
roughly equal, see Fig. \ref{4k}. Note that our results hold 
even for $\alpha\gg 1$, where the three-component complex is much more stable 
than the intermediate $AB$, $AC$, and $BC$ states.

\begin{figure}[htb]
\centering
\includegraphics[totalheight=1.5in]{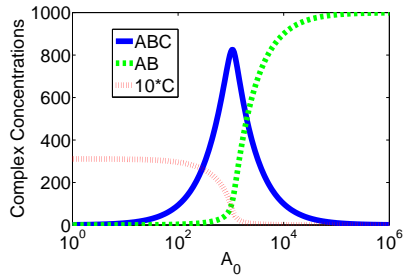}
\caption{(Color online) The dimensionless 
concentrations of the complex $ABC$ (solid line),
partial complex $AB$ (dashed line), and $C$ (dotted line) as a
function of the total concentration of $A$ particles, $A_0$ ($C$
is multiplied by 10 for visibility). $B_0$ and $C_0$ are fixed
$B_0 = C_0 = 10^3$. The maximum of $ABC$ for finite $A_0$ is a
result of the balance between increase in the number of $AB$ and
$AC$ complexes and the decrease in the number of available free
$B$ and $C$ particles as $A_0$ increases.} \label{ABCAB}
\end{figure}

We have explicitly checked that the same picture holds for
4-component complexes as well: fixing the concentrations $B_0$,
$C_0$, and $D_0$, the concentration of the target complex $ABCD$
is again maximized for $A_{0,max} \approx\max{(B_0,C_0,D_0)}$.
This behavior is expected to hold qualitatively for a general
number of components and arbitrary reaction rates, due to the
following argument: Assume a complex is to be produced from many
constituents, one of which ($A$) is far more abundant than the
others ($B$, $C$, ...). Since $A$ is in excess, almost all $B$
particles will bound to $A$ and form $AB$ complexes. Similarly,
almost all $C$ particles will bound to $A$ to form an $AC$
complex. Thus, there will be very few free $C$ particles to bound
to the $AB$ complexes, and very few free $B$ particles available
for binding with the $AC$ complexes. As a result, one gets
relatively many half-done $AB$ and $AC$ complexes, but not the
desired $ABC$ (note that $AB$ and $AC$ cannot bound together).
Lowering the concentration of $A$ particles allows more $B$ and
$C$ particles to remain in an unbounded state, and thus {\it
increases} the total production rate of $ABC$ complexes (Fig.
\ref{ABCAB}).

Many proteins take part in more than one complex. One might thus wonder 
what is the optimal concentration for these, and how it affects 
the general correlation observed between the concentrations of 
members of the same complex. In order to clarify this issue,
we have studied a model in which four proteins $A$, $B$, $C$ and $D$
bind together to form two desired products: the $ABC$ and $BCD$ complexes.
$A$ and $D$ do not interact, so that there are no complexes or
sub-complexes of the type $AD$, $ABD$, $ACD$ and $ABCD$. Solution of this 
model (see appendix) reveals that the efficiency of the production of $ABC$
and $BCD$ is maximized when (for a fixed ratio of $A_0$ and $D_0$)
$A_0+D_0\approx B_0\approx C_0$. One thus sees, as could have been expected, 
that proteins that are
involved in more than one complex 
(like $B$ and $C$ in the above model) will tend to have higher concentrations
than other members of the same complex participating in only one complex.
Nevertheless, since the protein-protein interaction network is scale-free,
most proteins take part in a small-number of complexes, and only a very
small fraction participate in many complexes. Moreover, given the three
orders of magnitude spread in protein concentrations (see figure 
\ref{lognormal}), 
only proteins participating in a very large number of complexes (relative to the avregae participation) or participating in two complexes of a very different
concentrations (i.e., $A_0\gg D_0$) will result in order-of-magnitude
deviations from the equal concentration optimum. 
The effects of these relatively
few proteins on the average over all interacting proteins
is small enough not to destroy the concentration correlation, as we observed
in the experimental data. 

In summary, the solution of our simplified complex formation model
shows that the rate and efficiency of complex formation depends
strongly, and in a non-obvious way, on the relative concentrations
of the constituents of the complex. The efficiency is maximized when all
concentrations of the different complex constituents are roughly
equal. Adding more of the ingredients beyond this optimal point
not only reduces the efficiency, but also results in lower product
yield. This unexpected behavior is qualitatively explained by a
simple argument, and is expected to hold generally. Therefore,
effective formation of complexes in a network puts constraints on
the concentrations on the underlying building blocks. Accordingly,
one can understand the tendency of members of cellular
protein-complexes to have uniform concentrations, as presented in
the previous Section, as a selection towards efficiency.

\appendix*
\section{Two coupled complexes}
We consider a model in which four proteins $A$, $B$, $C$ and $D$
bind together to form two desired products: the $ABC$ and $BCD$ complexes.
$A$ and $D$ do not interact, so that there are no complexes or
sub-complexes of the type $AD$, $ABD$, $ACD$ and $ABCD$. 
For simplicity, we assume the totally symmetric situation, 
where all the ratios of association
coefficients to their corresponding dissociation coefficients are equal,
i.e., the ratios $k_{d_{x,y}}/k_{a_{x,y}}$ are all equal to
$X_0$ and $k_{d_{x,y,z}}/k_{a_{x,y,z}}=X_0^2$, where $X_0$ is a constant with 
concentrations units. The extension
to the more general case discussed in the paper is straight forward.
Using the same scaling
as above, the reaction equations are solved by the substitutions $AB = A \cdot
B$, $AC = A \cdot C$, $BC = B \cdot C$, $BD=B\cdot D$, $CD=C\cdot D$,
$ABC = A \cdot B \cdot C$, and $BCD=B\cdot C\cdot D$, 
and one needs only to solve the material conservation
equations, which take the form:

\begin{eqnarray}
\label{eqA}
&A& + A \cdot B + A \cdot C + A \cdot B \cdot C = A_0\\
&B& + A \cdot B + B \cdot C + B \cdot D + A \cdot B \cdot C + B \cdot C \cdot D = B_0\nonumber\\ \label{eqB}\\
&C& + A \cdot C + B \cdot C + C \cdot D + A \cdot B \cdot C + B \cdot C \cdot D = C_0\nonumber\\ \label{eqC} \\
\label{eqD}
&D& + B \cdot D + C \cdot D + B \cdot C \cdot D = D_0
\end{eqnarray}

Denoting $\gamma \equiv \frac{D_0}{A_0}, D' \equiv \frac{D}{\gamma}$, 
Eq (\ref{eqD}) becomes 
\begin{equation}
D' + D' \cdot B + D' \cdot C + D' \cdot B \cdot C = A_0
\end{equation}
This is exactly the equation we wrote for A (\ref{eqA}), and thus 
$D = \gamma A$.
Substitutng this into equations (\ref{eqB}) and (\ref{eqC}), one gets
\begin{eqnarray}
\label{newB}
B + B \cdot C + (\gamma + 1)A \cdot B + (\gamma + 1)A \cdot B \cdot C = B_0\\
\label{newC}
C + B \cdot C + (\gamma + 1)A \cdot C + (\gamma+1)A \cdot B \cdot C = C_0
\end{eqnarray}
We now define $A' \equiv (\gamma + 1)A$, $A'_0 \equiv(\gamma+1)A_0$ and obtain
from (\ref{eqA},\ref{newB},\ref{newC})

\begin{eqnarray}
A' + A' \cdot B + A' \cdot C + A' \cdot B \cdot C &=& A'_0\\
B + A' \cdot B + B \cdot C + A' \cdot B \cdot C &=& B_0\\
C + A' \cdot C + B \cdot C + A' \cdot B \cdot C &=& C_0
\end{eqnarray}

These are the very same equations that we wrote for the 3-particles
case where the desired product was $ABC$. Their solution showed that
efficiency is maximized at $A_0 \approx B_0 \approx C_0$. We thus
conclude that in the present 4-component scenario, the efficiency of
$ABC$ and $BCD$ (for fixed $\gamma$) is maximized when 
$(A_0+D_0)\approx B_0\approx C_0$.

\acknowledgements{
We thank Ehud Schreiber for critical reading of the manuscript and
many helpful comments.
E.E. is supported by an Alon fellowship at Tel-Aviv University.}

\end{document}